\documentclass[pdflatex,sn-mathphys,Numbered]{sn-jnl}

\usepackage{graphicx}%
\usepackage{multirow}%
\usepackage{amsmath,amssymb,amsfonts}%
\usepackage{amsthm}%
\usepackage{mathrsfs}%
\usepackage[title]{appendix}%
\usepackage{xcolor}%
\usepackage{textcomp}%
\usepackage{manyfoot}%
\usepackage{booktabs}%
\usepackage{algorithm}%
\usepackage{algorithmicx}%
\usepackage{algpseudocode}%
\usepackage{listings}%
\usepackage{hyperref}
\usepackage{subcaption}
\usepackage{todonotes}
\usepackage{tabularx}
\usepackage{array}

\theoremstyle{thmstyleone}%

\theoremstyle{thmstyletwo}%

\theoremstyle{thmstylethree}%

\begin{document}

\title{Predicting Survivability of Cancer Patients with Metastatic Patterns Using Explainable AI}

\author[1]{\fnm{Polycarp} \sur{Nalela}}\email{polycarpnalela@missouri.edu}
\author[3]{\fnm{Deepthi} \sur{Rao}}\email{raods@health.missouri.edu}
\author*[1,2]{\fnm{Praveen} \sur{Rao}}\email{praveen.rao@missouri.edu
}

\affil[1]{\orgdiv{Institute for Data Science and Informatics}, \orgname{University of Missouri}, \orgaddress{\street{22 Heinkel Building}, \city{Columbia}, \postcode{65201}, \state{MO}, \country{Country}}}

\affil*[2]{\orgdiv{Electrical Engineering and Computer Science}, \orgname{University of Missouri}, \orgaddress{\street{225 Naka Hall}, \city{Columbia}, \postcode{65201}, \state{MO}, \country{USA}}}

\affil[3]{\orgdiv{Pathology and Anatomical Sciences}, \orgname{MU University Hospital}, \orgaddress{\street{1 Hospital Dr}, \city{Columbia}, \postcode{65201}, \state{MO}, \country{Country}}}

\abstract{Cancer remains a leading global health challenge and a major cause of mortality. This study leverages machine learning (ML) to predict the survivability of cancer patients with metastatic patterns using the comprehensive MSK-MET dataset, which includes genomic and clinical data from 25,775 patients across 27 cancer types. We evaluated five ML models-XGBoost, Naïve Bayes, Decision Tree, Logistic Regression, and Random Fores using hyperparameter tuning and grid search. XGBoost emerged as the best performer with an area under the curve (AUC) of 0.82. To enhance model interpretability, SHapley Additive exPlanations (SHAP) were applied, revealing key predictors such as metastatic site count, tumor mutation burden, fraction of genome altered, and organ-specific metastases. Further survival analysis using Kaplan-Meier curves, Cox Proportional Hazards models, and XGBoost Survival Analysis identified significant predictors of patient outcomes, offering actionable insights for clinicians. These findings could aid in personalized prognosis and treatment planning, ultimately improving patient care.}

\keywords{Metastatic Cancer, Machine Learning, Explainable AI}

\maketitle

\section{Introduction}
\label{sec-introduction}

The development of cancer is a complex process that occurs when genetic and epigenetic changes accumulate in the deoxyribose nucleic acid (DNA) of a cell. This leads to uncontrolled cell growth and invasion, which can ultimately result in the formation of a tumor. To better understand this disease and improve patient outcomes, researchers have traditionally relied on statistical and computational methods to analyse large datasets containing genomic, proteomic, and clinical information. However, with the emergence of artificial intelligence (AI) and ML, scientists are now able to develop more sophisticated models that can uncover patterns and features within these datasets, providing new insights into cancer biology, diagnosis, prognosis, treatment, and outcomes. 

ML models have achieved good success in recent years by outperforming traditional statistical models in cancer diagnosis and prognosis. In fact, a deep learning algorithm developed by Esteva et al.\cite{Esteva2017} achieved a sensitivity of 97\% and a specificity of 78\% in classifying skin lesions as benign or malignant. Later, Liu et al.~\cite{Liu2018} created a ML model that predicted the risk of lung cancer using computed tomography (CT) images, achieving an AUC-ROC score of 0.94. Other studies have also demonstrated the accuracy of ML in predicting breast cancer survival~\cite{Montazeri2016}, lymph node metastasis~\cite{Zhou2020}, breast cancer risk factors~\cite{Nindrea2018}, soft tissue sarcoma diagnosis and survival prediction\cite{Foersch2021}, lung cancer patient survival~\cite{Doppalapudi2021}, and prostate cancer~\cite{Zhang2021, Huang2021}. These findings highlight the potential of ML in improving cancer diagnosis and prognosis. One major advantage of supervised ML models is their ability to learn from large datasets of labeled examples, allowing for better generalization performance. For instance, Zhou et al.~\cite{Zhou2020} developed an ML model to predict the risk of breast cancer using mammography images. With over 200,000 mammography images and clinical data from over 60,000 patients, the model achieved an AUC score of 0.84 on a validation dataset. This demonstrates the potential of utilizing big data and ML in cancer diagnosis. 

Biomarkers play a crucial role in early detection, diagnosis, prognosis, and monitoring of cancer. Traditional biomarker discovery methods have some limitations, such as low sensitivity, specificity, and reproducibility. Additionally, they require prior biological knowledge or hypotheses. However, ML has emerged as a promising approach for cancer biomarker discovery, as it can integrate multiple data types and identify complex relationships between features and outcomes. Several ML algorithms, including Random Forests\cite{breiman2001random}, Support Vector Machines~\cite{cortes1995support}, and neural networks\cite{lecun2015deep} have been successfully applied to cancer biomarker discovery. These algorithms have been used to identify biomarkers from diverse data types such as gene expression~\cite{Liu2022, He2020, Huang2018}, micro ribonucleic acid (microRNA/miRNA) expression~\cite{Nguyen2022, Stefanou2022, Pawelka2022, Ghobadi2022}, DNA methylation~\cite{Ding2019, Liang2021}, miRNAs~\cite{Zhao2021}, and imaging data~\cite{Avanzo2020, Janssen2022, Stanzione2021, Wang2022}. 

In 2022, Nguyen et al.~\cite{Nguyen2022b} conducted genomic characterization of metastatic patterns in a large cohort of cancer patients. They published a dataset called MSK-MET containing a pan-cancer cohort of tumor genomic and clinical outcome data from 25,775 cancer patients. MSK-MET is a reliable data source containing information from extensive studies that were conducted on different cancer types from patients across different locations, demographics, and varying clinical features. Nguyen et al. ~\cite{Nguyen2022b} focused on genomic analysis and applying traditional statistical techniques to understand how mutation burden correlates with cancer outcomes. For example, they showed that tumor mutation burden strongly correlated with metastasis in cancer patients. However, they did not explore how ML can be employed to gain deeper insights from their data.

In this paper, we investigate how state-of-the-art ML techniques can be effective in predicting the survival status of cancer patients based on demographics, metastatic patterns, and clinical outcomes from MSK-MET. We further elaborate on the explainability of the best model using SHAP ~\cite{Lundberg2017}.

We believe that our investigation using the state-of-the-art ML techniques can be effective in predicting the survival status of cancer patients based on demographics, metastatic patterns, and clinical outcomes from MSK-MET. The key contributions of this work are as follows: 

\begin{itemize}
\item The study's application of the XGBoost classifier with a high AUC-ROC score of 0.82 offers a reliable tool for clinicians to assess patient prognosis with greater accuracy. This can lead to early and accurate identification of high-risk patients, enabling timely interventions and personalized treatment plans. 

\item The use of SHAP to elucidate key predictive factors such as metastases count, tumor mutation burden, and fraction of altered genome provides clinicians with actionable insights into the most critical determinants of patient outcomes. This knowledge aids in more informed decision-making regarding treatment options and resource allocation.

\item Ultimately, by improving the accuracy and transparency of survival prediction, this work aims to facilitate the development of tailored therapeutic strategies, enhances patient care, and potentially reduces healthcare costs through better management of metastatic cancer cases.
\end{itemize}




\section{Methodology}
\label{sec-method}

The figure \ref{fig:methodology} below illustrates steps that were followed for predicting cancer survivability using explainable AI. Row data was used to initially train ML models followed by SHAP analysis. Top features identified by SHAP were then further used in the survival analysis. The subsequent steps below detial how this was implemented.

\begin{figure}
    \centering
    \includegraphics[width=5.0in]{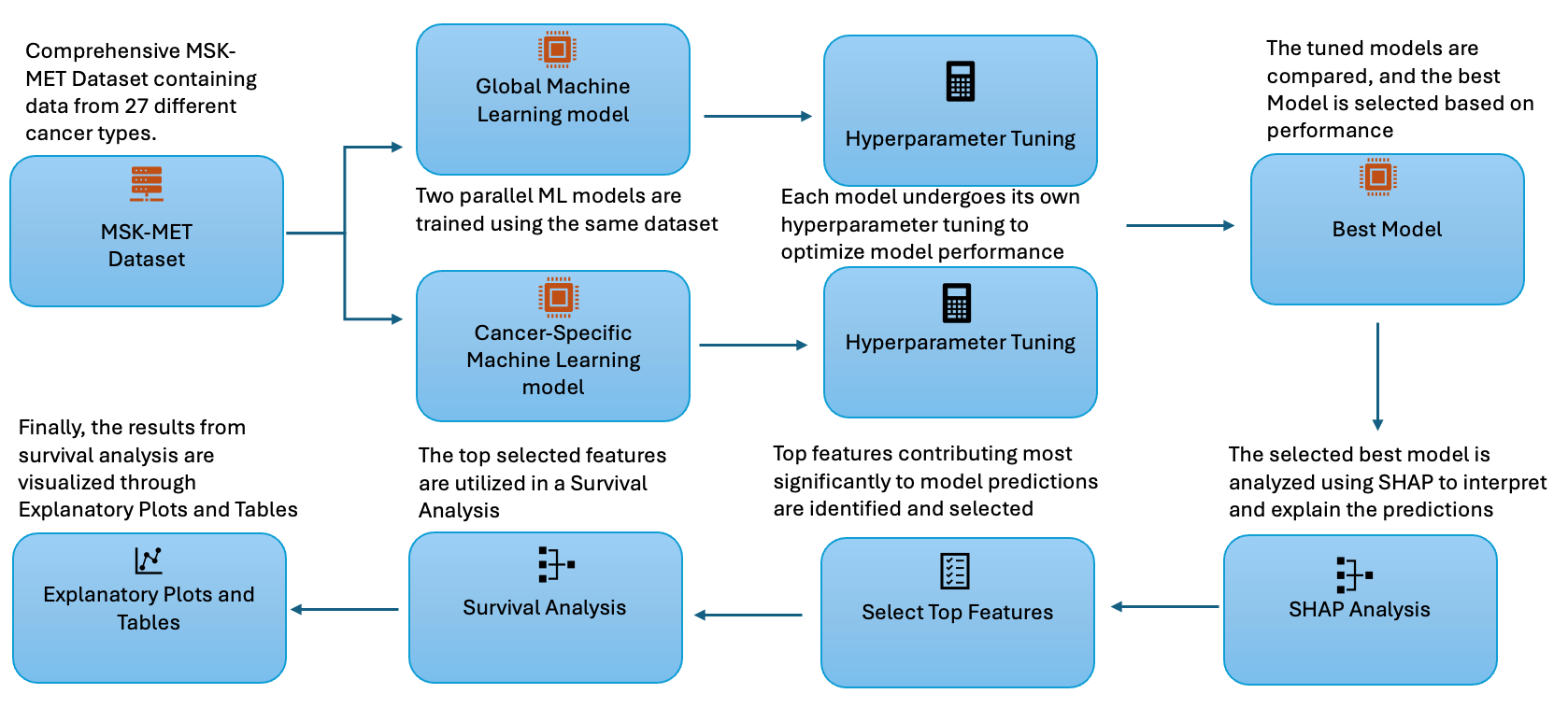}
    \caption{An abstract implementation of the ML models for predicting metastatic cancer survivability}
    \label{fig:methodology}
\end{figure}

\subsection{Data pre-processing}
We first performed a thorough exploratory data analysis (EDA) on MSK-MET that contained information from 25,775 cancer patients. Our EDA process began with a comprehensive analysis of  the dataset including the distribution of different cancer types, stages, and other relevant features. This helped us gain a deeper understanding of the underlying patterns and structures, which informed subsequent pre-processing steps. We pre-processed the input dataset and dropped columns (such as patient ID) and rows having large proportions of missing data. The final set contained 20,338 patients with 39 variables for each patient. In total, there were 27 cancer types. (See Table~\ref{table:cancer_frequency_count} for more details.) The overall survival status was the target variable for prediction. Categorical variables were encoded using label encoding, and features were scaled using Min-Max scaling to ensure that variables with larger magnitudes did not unduly influence model outcomes. The resulting pre-processed data was then split into training and testing sets for further analysis.

\begin{table}[htb]
    \caption{Frequency count of various cancer types}
    \centering
    \begin{small}
    \begin{tabular}{lc}
        \textbf{Cancer Type} & \textbf{Frequency Count} \\
        \hline
        Non-Small Cell Lung Cancer & 3,790 \\
        Colorectal Cancer & 2,696 \\
        Breast Cancer & 2,043 \\
        Pancreatic Cancer & 1,738 \\
        Prostate Cancer & 1,596 \\
        Endometrial Cancer & 988 \\
        Ovarian Cancer & 923 \\
        Melanoma & 882 \\
        Bladder Cancer & 870 \\
        Hepatobiliary Cancer & 790 \\
        Esophagogastric Cancer & 738 \\
        Soft Tissue Sarcoma & 420 \\
        Head and Neck Cancer & 362 \\
        Thyroid Cancer & 319 \\
        Renal Cell Carcinoma & 318 \\
        Gastrointestinal Stromal Tumor & 286 \\
        Small Cell Lung Cancer & 277 \\
        Germ Cell Tumor & 241 \\
        Mesothelioma & 219 \\
        Appendiceal Cancer & 160 \\
        Uterine Sarcoma & 133 \\
        Salivary Gland Cancer & 123 \\
        Gastrointestinal Neuroendocrine Tumor & 115 \\
        Skin Cancer, Non-Melanoma & 87 \\
        Cervical Cancer & 80 \\
        Small Bowel Cancer & 76 \\
        Anal Cancer & 68 \\
    \end{tabular}
    \end{small}
    \label{table:cancer_frequency_count}
\end{table}

\subsection{Stratified random sampling of training and testing sets}
We utilized a stratified random sampling approach to create the training and test sets. Firstly, we randomized the complete dataset to eliminate any inherent order or sequence. Then, we implemented stratification to ensure that the distribution of specific cancer types or stages in our training and testing sets mirrored that of the entire dataset. This is paramount to avoid potential biases and to ensure that our models have a representative sample of the different cancer types and stages present in the entire dataset. Following stratification, we allocated 80\% of the data (16,270 patient records) to the training set while reserving the remaining 20\% (4,068 patient records) for the test set. This approach provides a robust foundation for model development and validation, ensuring both broad and deep representation of the dataset in our training and testing phases.

\subsection{Selection of machine learning models}
This study used five machine learning algorithms XGBoost~\cite{chen2016xgboost}, Naïve Bayes, Decision Tree, Logistic Regression, and Random Forest to predict cancer survival rates using the MSK-MET dataset. XGBoost was selected for its efficiency in handling sparse data and combining models to improve accuracy through ensemble learning. Naïve Bayes, a simple classifier applying Bayes' theorem, was chosen for its efficiency in high-dimensional datasets. The Decision Tree, known for its easy visualization and handling of non-linear relationships, was included for its interpretability. Logistic Regression was utilized for binary classification, predicting survival probabilities, while Random Forest, an ensemble method using multiple decision trees, was chosen for its accuracy and control over over-fitting in large datasets. 

These algorithms were selected for their complementary strengths in addressing the dataset's high dimensionality, sparsity, and non-linearity, providing a well-rounded approach to predicting cancer survival rates.

\subsection{Grid-search with hyperparameter tuning}
Grid-search with hyperparameter tuning was applied to all five ML models. For XGBoost, parameters like n\_estimators (50-1000), max\_depth (1-10), and learning\_rate (0.01-0.3) were adjusted to optimize the number of trees, tree depth, and learning speed. Naïve Bayes was tuned by varying alpha (0.01-10.0), binarize (0.0, 0.5, 1.0), and fit\_prior (True/False). The Decision Tree's grid search adjusted max\_depth (None-10), min\_samples\_split, min\_samples\_leaf, and criterion ('gini' or 'entropy'). Logistic Regression was optimized with 'C' (0.001-1000), 'penalty' ('l1', 'l2', 'elasticnet', 'none'), and 'solver' ('newton-cg', 'lbfgs', 'liblinear', 'sag', 'saga'). Random Forest explored n\_estimators (50, 100, 200) and max\_features ('auto', 'sqrt').

Each model's performance was evaluated with 5-fold cross-validation, splitting the data into five parts for training and testing. This method ensures the best hyperparameters are selected, balancing model complexity and generalization, enhancing predictive accuracy for cancer survival rates in the MSK-MET dataset.

\subsection{Model architecture}
We initially applied a stacked ensemble ML approach to analyze the MSK-MET dataset, aiming to create a robust and explainable predictive model for diverse metastatic patterns.

The ensemble combined four base models Naïve Bayes, Decision Tree, Logistic Regression, and Random Forest chosen for their varied algorithms to capture different data patterns. Each model underwent hyperparameter tuning using grid search and cross-validation to ensure accuracy and generalizability.

In the stacking framework, these base models fed into an XGBoost model as the final estimator, leveraging gradient boosting to refine predictions and improve accuracy.

Building on this, we focused on enhancing the explainability of XGBoost, which had shown superior performance. Two strategies were employed: first, we created a unified XGBoost model without the 'cancer type' variable, analyzing clinical and demographic features across cancer types to find consistent predictive patterns. Second, we developed individual XGBoost models for the top five cancer types Non-Small Cell Lung Cancer, Colorectal Cancer, Breast Cancer, Pancreatic Cancer, and Prostate Cancer to capture unique interactions within each type, offering deeper insights into metastatic behaviors.

The purpose of employing these refined models was to delve deeper into the explainability of the predictive outcomes, thereby enabling physicians to make more informed and precise decisions. By understanding the predictive contributions of various features without the overarching category of cancer type, and by examining the distinct behaviours of individual cancer types, the models provided a dual perspective that combined a broad, generalized understanding with specific, detailed insights. This comprehensive approach enhanced the utility of the predictive models in clinical settings, where tailored, accurate predictions are crucial for effective patient care and treatment planning.

\subsection{Evaluation of the models}
After training and testing the ML models (XGBoost, Naïve Bayes, Decision Tree, Logistic Regression, and Random Forest) on the MSK-MET dataset, we assessed their performance using two key metrics: the classification report (Accuracy, Precision, Recall, F1-score, and Specificity) (Table~\ref{table:classifier_performance}) and the Area Under the ROC Curve (AUC-ROC) (Figure~\ref{fig:figureOne}). The AUC-ROC measures the model's ability to distinguish between classes, with higher AUC indicating better prediction. A score of 1 represents perfect predictions, 0.5 indicates random guessing, and below 0.5 suggests worse than random predictions. This is especially useful for imbalanced datasets, common in medical prognosis.

These metrics provide a comprehensive evaluation, ensuring the models not only predict accurately but also effectively identify positive cancer cases. This approach helps in selecting the best model for predicting cancer survival, balancing the need to detect true cases while minimizing false diagnoses. In the final evaluation, we focused on XGBoost, measuring only Accuracy and AUC-Score (Table~\ref{table:classifier_accuracy_auc}).

\subsection{Model interpretation and explanation}
To enhance the understandability and transparency of our predictive models, we utilized XGBoost and employed SHAP for model explainability. SHAP, based on game theory, provides a detailed and consistent measure of feature importance by computing each feature's contribution to the prediction. SHAP values represent a feature's responsibility for a change in the model output, ensuring local accuracy, missingness, and consistency. This method quantifies the impact of each feature on predictions and explains how the presence or absence of a feature affects the outcome. Beeswarm plots are particularly useful for visualizing SHAP values, showing features' influence and variability in a nuanced manner.

\subsection{Survival analysis}
Following the training of the XGBoost Machine Learning model and SHAP analysis, the most important features influencing patient survival were identified and used in the survival analysis. The primary goal was to examine the duration from cancer diagnosis to patient death, assessing how clinical and genomic variables impact survival times. We employed Kaplan-Meier Survival Analysis, Cox Proportional Hazards modeling, Log-Rank Tests for comparing survival distributions, and XGBoost Survival Analysis to deepen our understanding of patient outcomes. 

\subsubsection{Kaplan-Meier Survival Analysis}
The Kaplan-Meier estimator was utilized to evaluate survival probabilities over time across different patient subgroups. Patients were stratified based on key features identified from SHAP analysis, such as metastatic site count, tumor mutation burden, and specific organ metastases (e.g., liver, bone, lung). Survival curves were compared using the log-rank test to assess statistically significant differences between groups. A p-value of less than 0.05 was considered statistically significant.
\subsubsection{Cox Proportional Hazards Model}
The Cox PH model was applied to assess the influence of multiple covariates on patient survival while controlling for potential confounders. Key covariates included metastatic site count, fraction genome altered, tumor mutation burden, and distant metastases in specific organs. The proportional hazards assumption was evaluated using Schoenfeld residuals, and any violations were addressed through stratification or inclusion of time-varying covariates. Hazard ratios (HR) with corresponding 95\% confidence intervals (CI) were reported to quantify risk associations.

\subsubsection{Log-Rank Test}
To further compare survival distributions between different patient cohorts, the log-rank test was applied. This test was used to determine whether survival differences observed between patient subgroups (e.g., metastatic vs. non-metastatic) were statistically significant. The resulting p-values guided the identification of meaningful clinical predictors. Furthermore, a plot was generated for the Kaplan-Meier Survival Curves with the Overall Survival (Months) on the X-axis and Survival Probability on the Y-axis.

\subsubsection{XGBoost Survival Analysis}
To capture complex, non-linear relationships and interactions among variables, XGBoost Survival Analysis was implemented. This adaptation of XGBoost used a Cox loss function to accommodate censored survival data. Hyperparameter tuning was conducted using grid search, optimizing parameters such as n\_estimators, max\_depth, and learning\_rate. The model’s concordance index (C-index) was used to evaluate predictive performance. SHAP values were also applied to the survival model to interpret feature importance and explore individual risk predictions.

\section{Results}
\label{sec-results}

\begin{figure}
    \centering
    \includegraphics[width=3.0in]{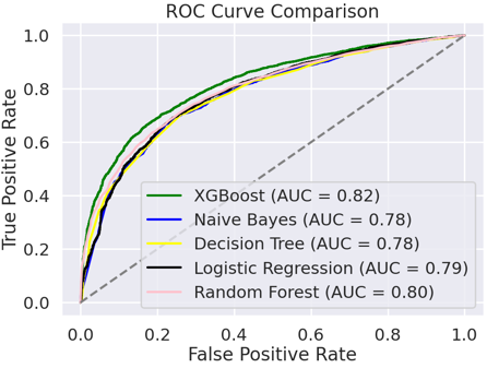}
    \caption{AUC comparison for the five ML models in the ensemble setup}
    \label{fig:figureOne}
\end{figure}

The evaluation of five distinct models on the MSK-MET dataset yielded a spectrum of performances, with each model's efficacy determined using a variety of metrics. Notably, each model's ability to distinguish between classes was assessed using  AUC-ROC scores.  The classification performance is reported in Table~\ref{table:classifier_performance} while the AUC-ROC curves are shown in Figure~\ref{fig:figureOne} and Table~\ref{table:classifier_accuracy_auc}.

\begin{table}[htb]
    \caption{Performance metrics for the ensemble classifier and individual models; best value shown in bold}
    \label{table:classifier_performance}

    \centering
    \begin{small}
    \begin{tabular}{p{0.8in}ccccc}
        \textbf{Classifier} & \textbf{Precision} & \textbf{Recall} & \textbf{F1-Score} & \textbf{Accuracy} \\
        \hline
        XGBoost & \textbf{0.74} & 0.74 & \textbf{0.74} & \textbf{0.74} \\
        Naïve Bayes & 0.72 & 0.72 & 0.72 & 0.72 \\
        Decision Tree & 0.72 & 0.72 & 0.72 & 0.72 \\
        Logistic Regression & 0.72 & 0.72 & 0.72 & 0.72 \\
        Random Forest & 0.70 & \textbf{0.77} & 0.73 & 0.72 \\
        Ensemble classifier & 0.71 & 0.76 & \textbf{0.74} & 0.73 \\
    \end{tabular}
    \end{small}
\end{table}

\begin{table}[htb]
\centering
\resizebox{\textwidth}{!}{%
\begin{tabular}{lrrrr}
\hline
\textbf{Covariate} & \textbf{HR (exp(coef))} & \textbf{95\% CI (Lower)} & \textbf{95\% CI (Upper)} & \textbf{p-value} \\
\hline
Metastatic patient   & 2.18 & 1.97 & 2.42 & $<$0.005 \\
Met Site Count       & 1.03 & 1.02 & 1.04 & $<$0.005 \\
TMB (nonsynonymous)  & 1.00 & 0.99 & 1.00 & $<$0.005 \\
Fraction Genome Altered & 1.32 & 1.19 & 1.46 & $<$0.005 \\
Sample Type          & 0.87 & 0.83 & 0.90 & $<$0.005 \\
Distant Mets: Liver  & 1.81 & 1.73 & 1.90 & $<$0.005 \\
Distant Mets: Bone   & 1.43 & 1.37 & 1.50 & $<$0.005 \\
Distant Mets: Lung   & 1.16 & 1.11 & 1.22 & $<$0.005 \\
\hline
\end{tabular}%
}
\caption{Condensed Cox Proportional Hazards model displaying hazard ratios, confidence intervals, and p-values for the most important risk factors}
\label{table:coxPH_condensed}
\end{table}

Table~\ref{table:classifier_performance} reports the performance metrics for both the ensemble classifier and each individual model; best values are highlighted. The XGBoost model demonstrated the highest accuracy among the standalone models, achieving an accuracy of 0.74 and an AUC of 0.82. This was marginally superior to the other models, which showed relatively consistent results; Naïve Bayes and Decision Tree both recorded accuracy scores of 0.72 (AUC=0.78), Logistic Regression at 0.72 (AUC=0.79), and Random Forest at 0.72 (AUC=0.80). When incorporated into an ensemble framework, the stacked model slightly outperformed Naïve Bayes, Decision Tree, Logistic Regression, and Random Forest, but its overall performance remained slightly below that of XGBoost. This indicates that while the ensemble leverages the strengths of the base models, XGBoost alone was the most effective for this classification task.

Furthermore, a global XGBoost model was compared to cancer-specific XGBoost models for the top five cancer types - Non-Small Cell Lung Cancer, Colorectal Cancer, Breast Cancer, Pancreatic Cancer, and Prostate Cancer (Table~\ref{table:classifier_accuracy_auc}). The Prostate Cancer–specific model stood out with the highest accuracy (0.84) and AUC (0.88). Meanwhile, Pancreatic Cancer posted a lower AUC of 0.68, reflecting greater challenges in classification for that subgroup. These variations highlight differences in predictive complexity and emphasize the importance of tailoring models to specific cancer contexts.

\begin{table}[htb]
    \caption{Classifier accuracy and AUC score comparison for the global XGBoost model and the cancer specific XGBoost models}
    \label{table:classifier_accuracy_auc}
    \centering
    \begin{small}
    \begin{tabular}{p{1.5in}ccc}
        \textbf{Classifier} & \textbf{Accuracy score} & \textbf{AUC Score} \\
        \hline
        Global model & 0.74 & 0.82 \\
        Non-Small Cell Lung Cancer-model & 0.71 & 0.79 \\
        Colorectal Cancer model & 0.73 & 0.81 \\
        Breast Cancer model & 0.76 & 0.85 \\
        Pancreatic Cancer model & 0.72 & 0.68 \\
        Prostate Cancer model & 0.84 & 0.88 \\
    \end{tabular}
    \end{small}

\end{table}

\begin{figure}[h!]
\centering
\includegraphics[width=3.0in]{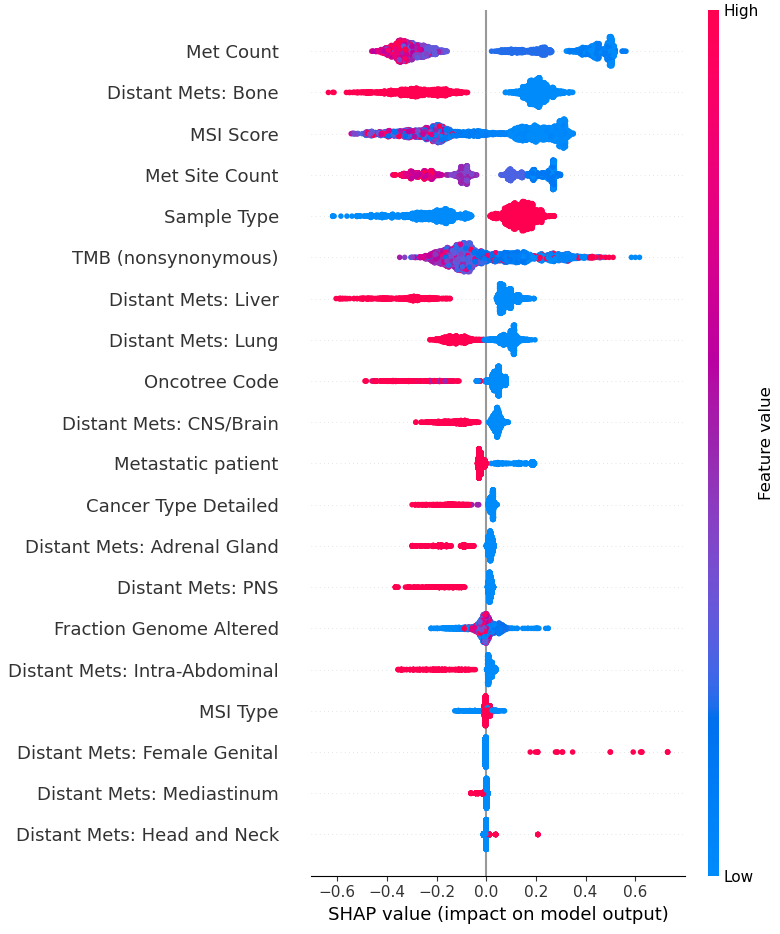}
\caption{SHAP-Beeswarm plot for Non Small Cell Lung Cancer}
\label{fig:shap-Non-Small Cell Lung Cancer}
\end{figure}

\begin{figure}[h!]
\centering
\includegraphics[width=3.0in]{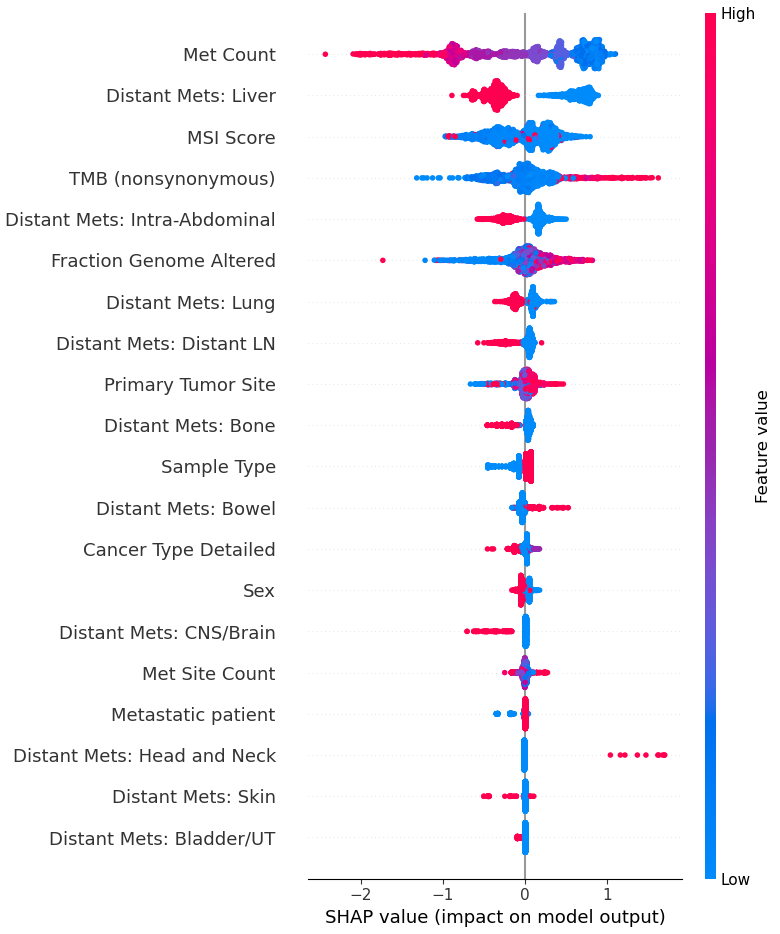}
\caption{SHAP-Beeswarm plot for Colorectal Cancer}
\label{fig:Colorectal Cancer}
\end{figure}

\begin{figure}[h!]
\centering
\includegraphics[width=3.0in]{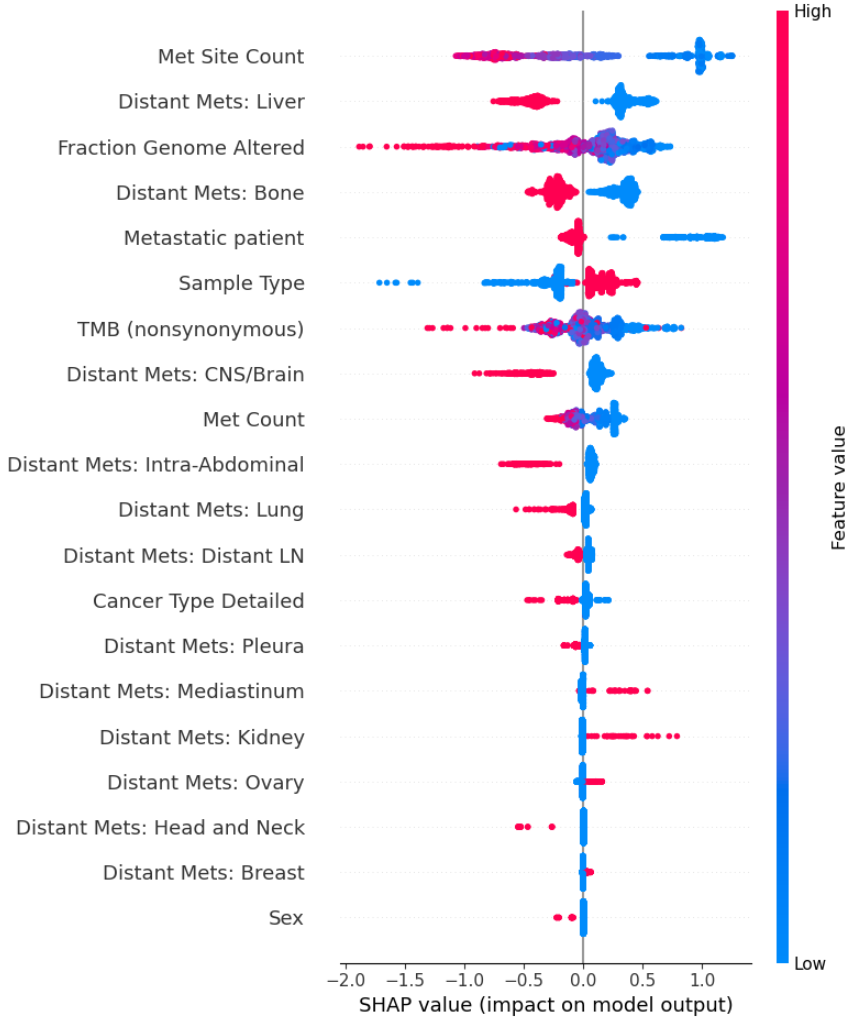}
\caption{SHAP-Beeswarm plot for Breast Cancer}
\label{fig:Breast Cancer}
\end{figure}

\begin{figure}[h!]
\centering
\includegraphics[width=3.0in]{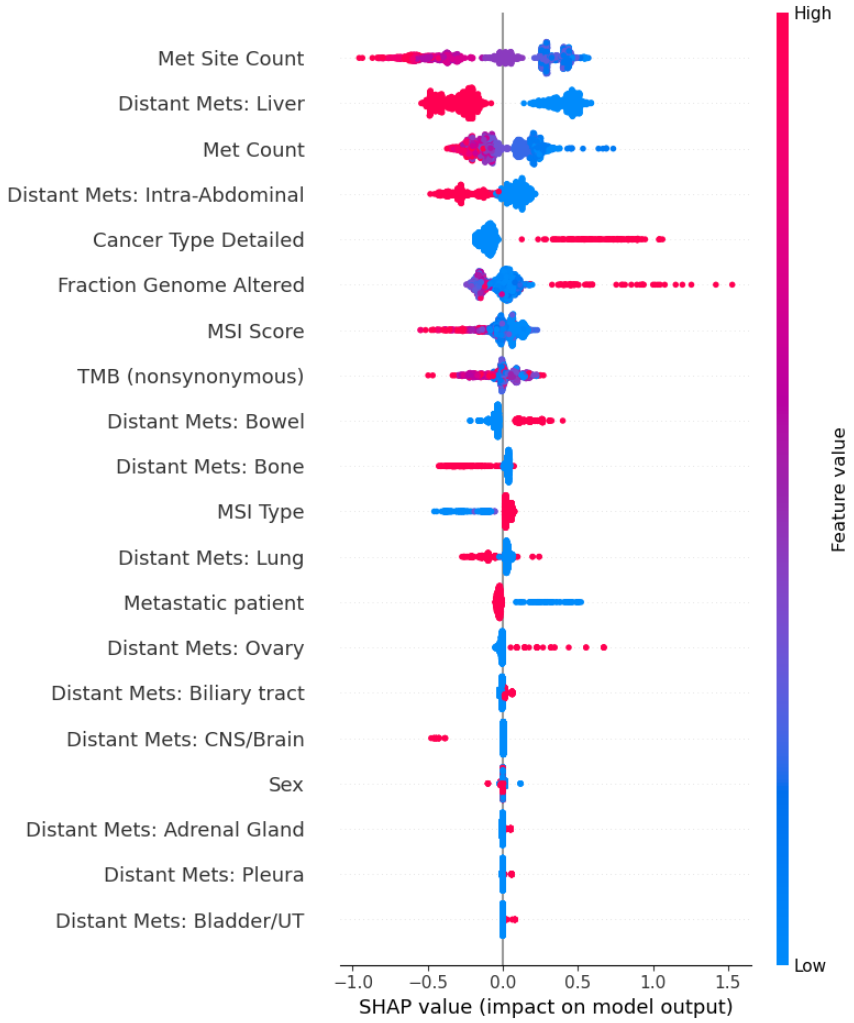}
\caption{SHAP-Beeswarm plot for Pancreatic Cancer}
\label{fig:Pancreatic Cancer}
\end{figure}

\begin{figure}[h!]
\centering
\includegraphics[width=3.0in]{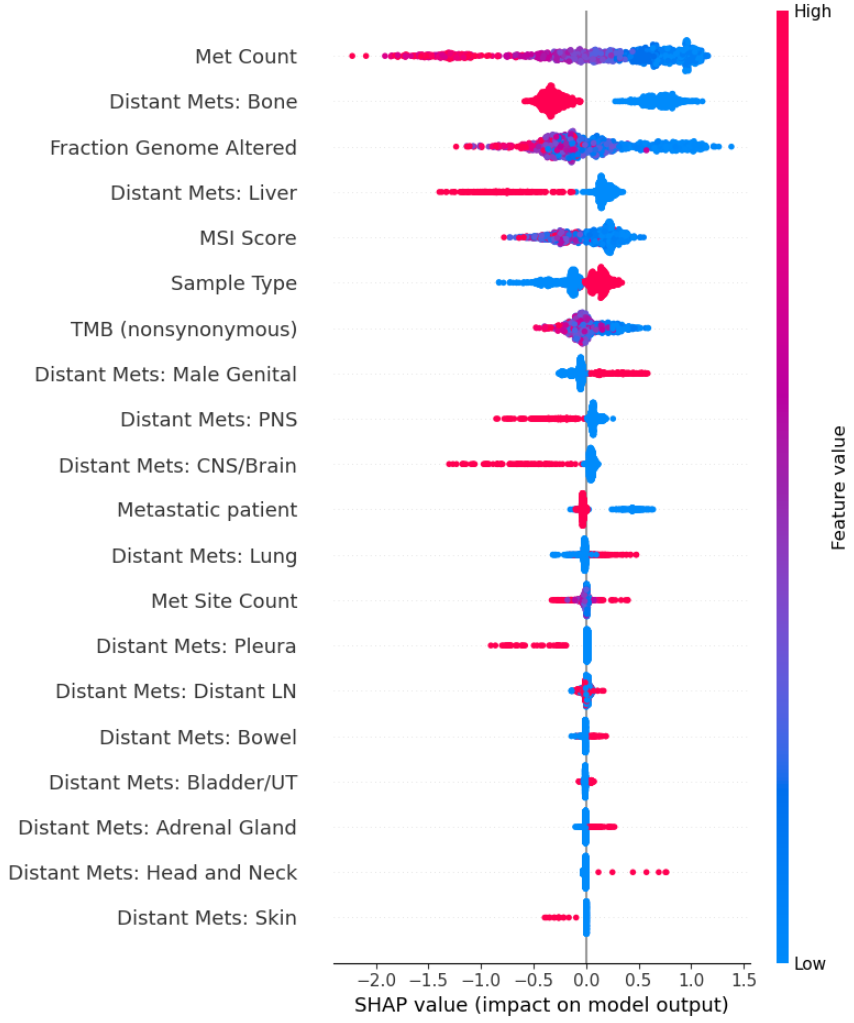}
\caption{SHAP-Beeswarm plot for Prostate Cancer}
\label{fig:Prostate Cancer}
\end{figure}

In the realm of explainability and feature impact, SHAP values were utilized to elucidate the contributions of different features to the XGBoost model. In our SHAP analysis, survival is the positive class. The analysis delineated a hierarchy of feature importance, with the top features contributing to model decisions including metastatic sites count, metastasis count, distant metastasis in the liver, tumor mutation burden, fraction of genome altered, and distant metastasis in the bone (Table~\ref{table:cancer_features}). These features were not only pivotal in terms of importance but also exhibited a notable distribution in the Beeswarm plots, where high and low values distinctly populated opposite sides, illustrating their robust predictive power (Figure~\ref{fig:shap-Non-Small Cell Lung Cancer}, \ref{fig:Colorectal Cancer}, \ref{fig:Breast Cancer}, \ref{fig:Pancreatic Cancer}, \ref{fig:Prostate Cancer}). 

\begin{figure}[h!]
\centering
\includegraphics[width=3.5in]{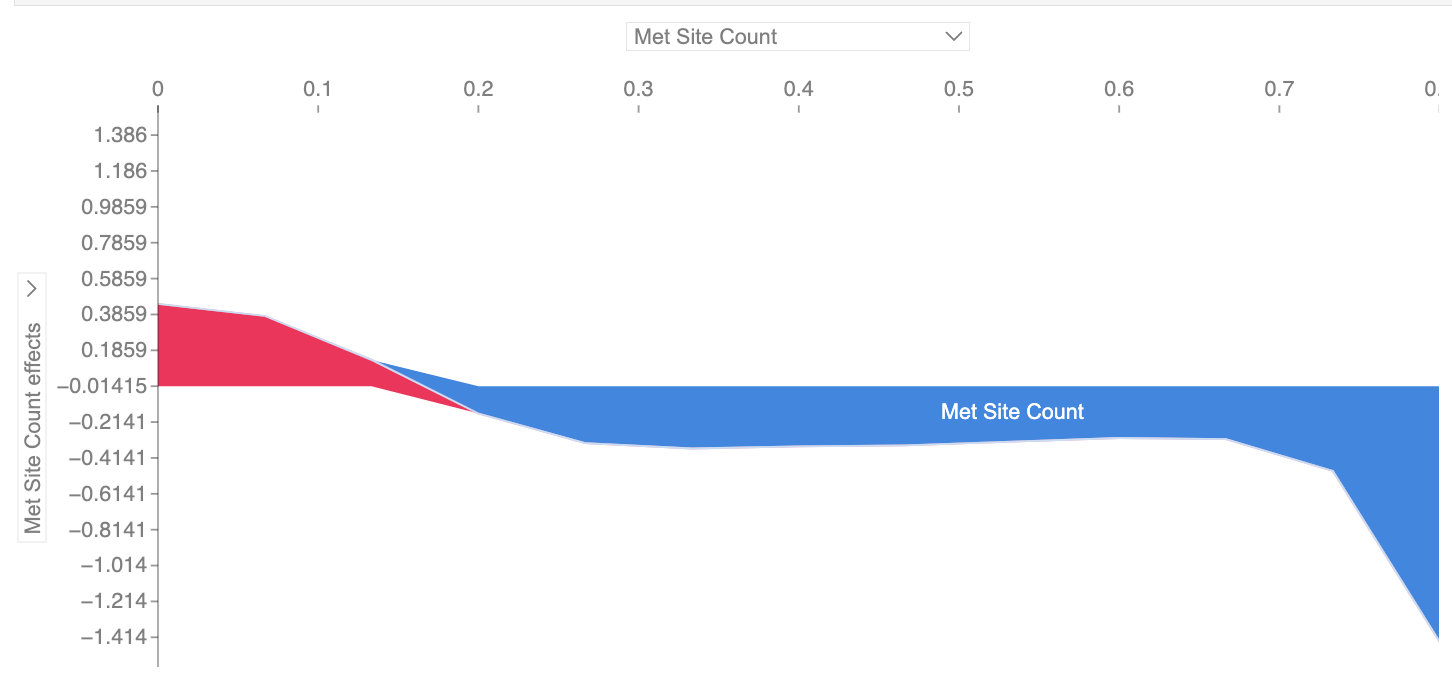}
\caption{SHAP-Force plot for met sites count for first 1000 observations}
\label{fig:forcePlot met sites count}
\end{figure}

Further insights were gleaned from the cancer-specific XGBoost models (Table \ref{table:cancer_features}, and Figure~\ref{fig:shap-Non-Small Cell Lung Cancer}, \ref{fig:Colorectal Cancer}, \ref{fig:Breast Cancer}, \ref{fig:Pancreatic Cancer}, \ref{fig:Prostate Cancer}, \ref{fig:forcePlot met sites count}), which underscored the importance of certain features in particular cancer contexts. For instance, Distant metastasis in the lung was a significant predictor in the model for Non-Small Cell Lung Cancer (Figure \ref{fig:shap-Non-Small Cell Lung Cancer}), Sample type emerged as a crucial feature in the breast cancer model (Figure \ref{fig:Breast Cancer}), and Distant metastasis in the male genital was notably influential in the prostate cancer model (Figure \ref{fig:Prostate Cancer}). This differentiation in feature importance across cancer types highlights the nuanced and tailored approach of the cancer-specific models, allowing for a more refined understanding of metastatic behaviors and their predictors.

The density and positioning of features such as metastatic sites counts and distant metastasis in the bone in SHAP plots underscored their strong influence on model predictions. For example, the force plots in Figure  \ref{fig:forcePlot met sites count} particularly illustrated the impact of increasing metastatic sites count, which corresponded with an increase in the likelihood of a positive prediction, thus reinforcing the predictive value of these features.

\begin{table}[htb]
    \caption{Cancer features and their ranks}
    \label{table:cancer_features}
    \centering
    \begin{small}
    \begin{tabularx}{\textwidth}{p{1in} X X X X X}
        \textbf{Model Type} & \textbf{Feature Rank 1} & \textbf{Feature Rank 2} & \textbf{Feature Rank 3} & \textbf{Feature Rank 4} & \textbf{Feature Rank 5} \\
        \hline
        Global model & Distant Mets: Liver & Met Site Count & Tumor Mutation Burden & Distant Mets: Bone & Fraction of Genome \\
        Non-Small Cell Lung Cancer & Met Site Count & Fraction of Genome Altered & Tumor Mutation Burden & Distant Mets: Lung & Distant Mets: Bone \\
        Colorectal Cancer & Met Count & Distant Mets: Liver & Tumor Mutation Burden & Distant Mets: Intra-abdominal & Fraction of Genome Altered \\
        Breast Cancer & Met Site Count & Distant Mets: Bone & Distant Mets: Liver & Fraction of Genome Altered & Sample Type \\
        Pancreatic Cancer & Distant Mets: Liver & Met Count & Fraction of Genome Altered & Met Site Count & Tumor Mutation Burden \\
        Prostate Cancer & Met Count & Distant Mets: Bone & Fraction of Genome Altered & Distant Mets: Liver & Distant Mets: Male Genital \\
    \end{tabularx}
    \end{small}
\end{table}




\begin{table*}[tbh]
\caption{Comparison of our paper and recent papers related to predicting survival of patients with metastatic cancer. GM and LM stand for Global model and Local model (Cancer-specific model), respectively.}
\label{table:papers_comparison}
\centering
\begin{small}
\begin{tabularx}{\textwidth}{>{\centering\arraybackslash}X >{\centering\arraybackslash}X >{\centering\arraybackslash}X >{\centering\arraybackslash}X >{\centering\arraybackslash}X >{\centering\arraybackslash}X >{\centering\arraybackslash}X >{\centering\arraybackslash}X >{\centering\arraybackslash}X}

\textbf{Paper} & 
\textbf{XGBoost } & 
\textbf{Naïve  Bayes } & 
\textbf{Decision Tree} & 
\textbf{Logistic Regression} & 
\textbf{Random Forest} & 
\textbf{GM/LM} & 
\textbf{SHAP Analysis} & 
\textbf{Survival Analysis} \\

\hline

Zhao et al. 2020 & $\times$ & $\times$ & $\times$ & $\times$ & $\times$ & $\times$ & $\times$ & $\times$ \\
Tapak et al. 2019 & $\times$ & $\checkmark$ & $\times$ & $\times$ & $\checkmark$ & $\times$ & $\times$ & $\times$ \\
Kourou et al. 2015 & $\times$ & $\times$ & $\checkmark$ & $\times$ & $\times$ & $\times$ & $\times$ & $\times$ \\
Nicolò et al. 2020 & $\times$ & $\times$ & $\times$ & $\times$ & $\checkmark$ & $\times$ & $\times$ & $\times$ \\
Maouche et al. 2023 & $\checkmark$ & $\times$ & $\checkmark$ & $\checkmark$ & $\times$ & $\times$ & $\checkmark$ & $\times$ \\
\textbf{Our Paper} & $\checkmark$ & $\checkmark$ & $\checkmark$ & $\checkmark$ & $\checkmark$ & $\checkmark$ & $\checkmark$ & $\times$ \\
\end{tabularx}
\end{small}
\end{table*}

\begin{figure}[h!]
\centering
\includegraphics[width=3.0in]{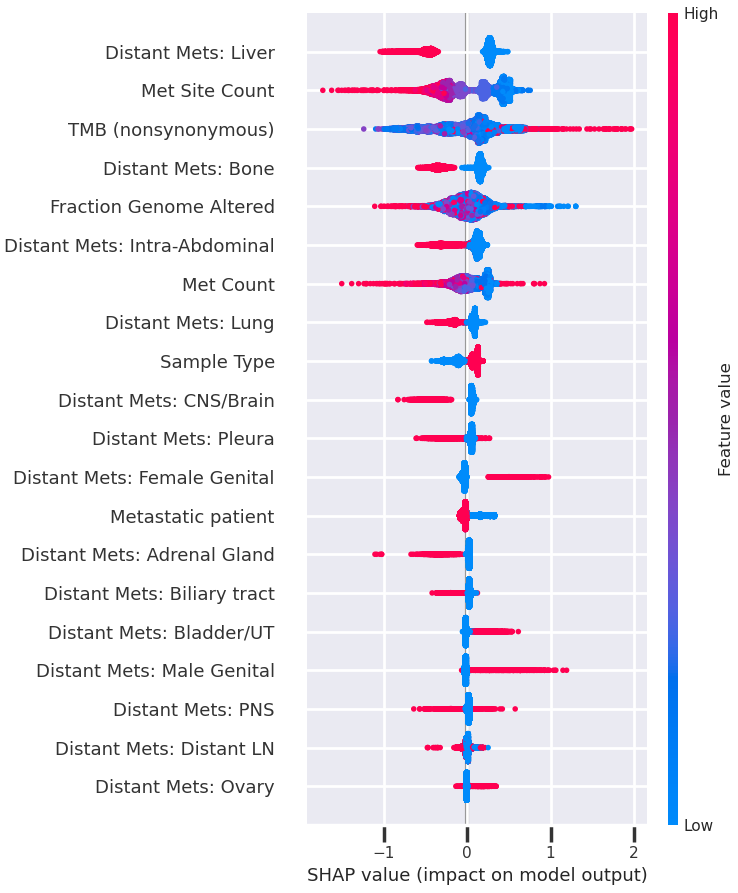}
\caption{SHAP-Beeswarm plot for Unified XGBoost model}
\label{fig:Unified XGBoost model}
\end{figure}

\begin{figure}
    \centering
    \includegraphics[width=5.0in]{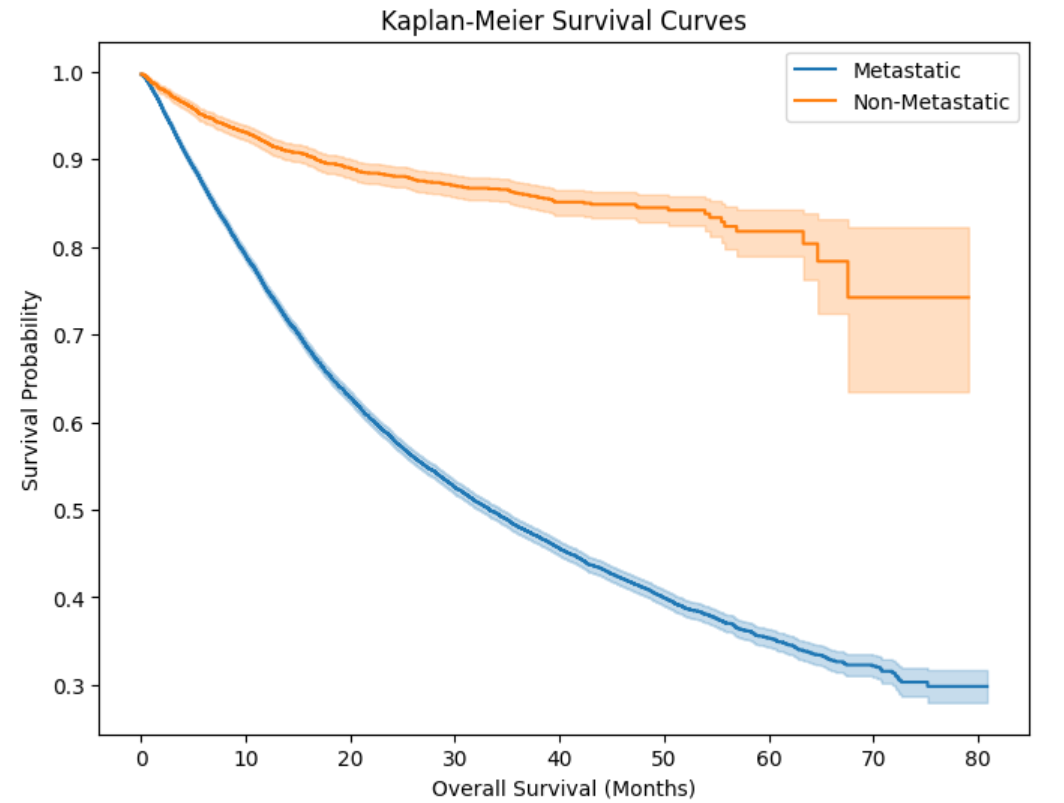}
    \caption{The keplan meier survival curves for metastatic and non-metastatic patients}
    \label{fig:kaplan}
\end{figure}

In the Kaplan-Meier analysis (Figure \ref{fig:kaplan}), patients were stratified into “Metastatic” and “Non-metastatic” groups to compare differences in overall survival. The survival probability of patients in the Metastatic group was notably lower than that of the Non-metastatic group, as seen in the pronounced separation of their survival curves. By approximately 80 months, patients with metastatic disease exhibited a survival probability of 0.3, whereas those without metastases had a corresponding survival probability of 0.8, underscoring the substantial impact of metastatic status on long-term survival outcomes.

In the Cox Proportional Hazards model (Table \ref{table:coxPH_condensed}), metastatic site count, tumor mutation burden, fraction of genome altered, and distant metastases (particularly in the liver and bone) displayed hazard ratios above 1.0, indicating an increased risk of mortality. These relationships attained statistical significance, with p-values under the established threshold. The proportional hazards assumption was checked through Schoenfeld residuals, and only minor deviations were noted, which did not substantially affect the covariate estimates. The model’s concordance index (C-index) reached approximately 0.66, reflecting moderate predictive power in distinguishing survival outcomes among different patient subgroups. 

An XGBoost survival model, fitted with a Cox-based loss function, was then used to incorporate non-linear effects and potential feature interactions more explicitly. After hyperparameter tuning, the XGBoost model achieved a higher C-index (0.7) than the standard Cox model. In Table \ref{table:Important_Features_of_the_XGBoost_Survival_Analysis}, the model’s  important features are displayed with metastasis, tumor mutation burden, fraction of genome altered among others listed as the most important features that influence the prediction.

\begin{table}[htb]
    \caption{Important Features of the XGBoost Survival Analysis}
    \centering
    \begin{small}
    \begin{tabular}{lr}
        \textbf{Feature} & \textbf{Importance} \\
        \hline
        Distant Mets: Liver & 0.315575 \\
        Fraction Genome Altered & 0.155434 \\
        TMB (nonsynonymous) & 0.138214 \\
        Met Site Count & 0.103948 \\
        Metastatic patient & 0.102230 \\
        Distant Mets: Bone & 0.092516 \\
        Distant Mets: Lung & 0.048726 \\
        Sample Type & 0.043358 \\
    \end{tabular}
    \end{small}
    \label{table:Important_Features_of_the_XGBoost_Survival_Analysis}
\end{table}

\section{Discussion}
\label{sec-discussion}

\subsection{Novelty of Our Work}
\label{sec-novelty}

Although significant studies~\cite{zhao2020predicting, tapak2019prediction, nicolo2020machine, kourou2015machine, maouche2023explainable} have been conducted in the realm of predicting cancer survivability, our work stands out by bringing new enhancements that significantly contribute to better prediction of cancer survivability, particularly by thorough comparison of machine learning models, the strategic use of both global and cancer-specific models, and in-depth model explainability using SHAP values.

Firstly, we begin by comparing five different machine learning models XGBoost, Naïve Bayes, Decision Tree, Logistic Regression, and Random Forest each rigorously tuned using exhaustive grid search for hyperparameters. This approach ensures that each model is optimized for the specific task of predicting cancer survivability, a detail that is often overlooked in existing literature. Many studies tend to focus on one or two models, without ensuring that the models are fully optimized for comparison. For example, prior work done by Zhao et al., 2020,~\cite{zhao2020predicting}, Tapak et al., 2019~\cite{tapak2019prediction} and Nicolò et al., 2020~\cite{nicolo2020machine} (Table~\ref{table:papers_comparison}), evaluate models but lack the thoroughness in hyperparameter tuning that our study provides. We believe that this rigorous approach enhances the reliability of our findings and provides a more comprehensive understanding of which model performs best under specific conditions.

Moreover, we employ a methodology that is designed to first use a global model to gain a general overview of the most important patterns and predictors for metastatic cancer survivability, followed by a deeper dive into cancer-specific models. This two-tiered approach is critical because it allows us to identify broad patterns while also uncovering nuances that might be missed or misinterpreted in a global model. Many published studies, such as Kourou et al., 2015~\cite{kourou2015machine} and Zhao et al., 2020~\cite{zhao2020predicting} (Table~\ref{table:papers_comparison}), predominantly dwell on global accuracy metrics without taking this crucial next step to explore more specific patterns within subgroups of the data. By contrast, our approach provides a dual perspective broad insights from the global model and detailed, cancer-specific insights that we believe are essential for advancing personalized medicine.

Finally, the use of SHAP values in our study is particularly noteworthy. We did not just stop at model performance but delved deep into explainability, first for the global model and then for the cancer-specific models. This process allowed us to generate refined explainability that highlights not just which features are important, but how their importance varies across different types of cancer. The use of SHAP in both global and specific contexts is a novel approach that we believe adds substantial value to the study. While many studies, such as Maouche et al., 2023~\cite{maouche2023explainable} (Table~\ref{table:papers_comparison}), employ SHAP or similar methods, they often do so at a surface level, without the comprehensive, model-specific analysis that we provide. This depth of analysis is crucial for understanding the true implications of the model’s predictions and for making informed clinical decisions.

\subsection{Explainability}

The present study aimed to elucidate explainability of ML towards predicting cancer survival. Our evaluation criteria were centered around two main performance metrics: accuracy and AUC-ROC. These metrics guided us to select XGBoost as the best model and used it for further analysis.

It is noteworthy that while accuracy provides a direct interpretation of model performance, the AUC-ROC score is a more nuanced metric, offering insight into the trade-off between the true positive rate and the false positive rate. A higher AUC-ROC value implies a better model performance overall, particularly in the domain of medical prognosis where the costs of false negatives are often high~\cite{fawcett2006introduction}.

The XGBoost model exhibited the highest performance among the individual models with an accuracy of 0.74 and an AUC of 0.82. This aligns with previous research indicating that XGBoost is highly effective for complex classification tasks due to its handling of varied data types and its capability of capturing non-linear relationships~\cite{chen2016xgboost}. The marginally lower performance of the other models (Naïve Bayes, Decision Tree, Logistic Regression, and Random Forest) with accuracies and AUCs around 0.72 and 0.78--0.80 respectively, suggests that while effective, these models may not fully capture the complexities of metastatic cancer data as effectively as gradient boosting methods.

The ensemble model, which slightly outperformed the standalone models of Naïve Bayes, Decision Tree, Logistic Regression and Random Forest but still lower than XGBoost (Table~\ref{table:classifier_performance}), supports the hypothesis that combining multiple learning algorithms can lead to improved prediction performance by leveraging the diverse strengths of various base models~\cite{wolpert1992stacked}. However, the modest increase in performance also indicates diminishing returns, possibly due to overlapping strengths among the chosen models. This phenomenon where ensemble gains are marginal has been observed in other medical applications and may reflect an intrinsic limitation in the diversity of modeling approaches applied~\cite{kuncheva2003measures}.

The SHAP analysis revealed critical insights into the features driving the models' predictions. Features such as Metastatic sites count, Metastasis count, and specific locations of distant metastasis (liver, bone) being identified as highly influential aligns with clinical understandings of cancer prognosis. Tumor mutation burden and Fraction of genome altered are well-documented predictors of cancer outcomes and their prominence in the model corroborates their biological and clinical relevance~\cite{davoli2017tumor}.

The differential importance of features across cancer-specific models such as the prominence of lung metastasis in Non-Small Cell Lung Cancer and Prostate Cancer as well as the significance of sample type in breast cancer highlights the tailored nature of these models. This specificity is crucial for clinical applications as it aligns with personalized medicine approaches, where treatment and prognosis are increasingly based on individual patient characteristics and cancer profiles~\cite{tsimberidou2015targeted}.

In clinical relevance, these outcomes have several implications for clinical practice and further research. Firstly, the ability of the ensemble model to slightly outperform individual models suggests that such approaches could be more widely adopted in clinical settings to improve the accuracy of metastasis predictions, which is critical for treatment planning and patient management.

Secondly, the insights from SHAP analysis facilitate a deeper understanding of which features are most predictive of metastasis, aiding in the identification of potential biomarkers for early detection and targeted therapy. For instance, the significance of specific metastatic sites could lead to more focused monitoring strategies for patients at high risk of metastasis to these locations.

Lastly, the feature ``Met Count'' mirrors ``Met Site Counts'' in its prognostic relevance but offers a nuanced perspective on the burden of metastasis, focusing on the lesion count. Its SHAP value distribution, leaning towards the positive, elucidates a direct correlation between the quantity of metastatic lesions and the model's inclination towards higher risk predictions. The visualization of this feature through color and density of SHAP values enriches our understanding of how the model quantifies and integrates the metastatic burden into its predictive framework.

\subsection{Survival Insights and Clinical Implications}

The survival analysis results reinforce and extend these classification insights by highlighting how metastatic status and specific genomic variables shape both risk level and the time horizon for adverse events. Kaplan-Meier curves showed that patients with metastatic disease had notably lower survival probabilities than non-metastatic patients, with a substantial gap (0.3 vs. 0.8) at approximately 80\,months. This large difference aligns with the SHAP-based identification of metastasis-related features as top predictors.

Cox Proportional Hazards modeling further quantified the impact of these predictors. Metastatic site count, fraction of genome altered, and tumor mutation burden exhibited hazard ratios above 1.0, indicating an elevated risk of mortality. The Cox model's concordance index, around 0.66, confirmed moderate discrimination between risk groups, though some non-linear interactions remained underrepresented.

Adapting XGBoost for survival data addressed these non-linearities to a greater extent. The XGBoost survival model improved upon the baseline Cox model, reaching a higher concordance index (0.700). This finding underlines the effectiveness of ensemble-based approaches in handling the complex interplay between clinical factors and genomic alterations.

From a clinical standpoint, these survival observations suggest that patients with high metastatic burden particularly in the liver and bone or those harboring substantial genomic alterations may require more aggressive or tailored interventions. By integrating classification and survival analyses, our approach enables both early identification of high-risk cases and a better understanding of when negative outcomes are likely to occur. In future work, additional data modalities (e.g., proteomics, imaging), as well as time-varying covariates, could be incorporated to further refine these survival models. Prospective clinical validation would also strengthen the real-world applicability of these findings, ensuring that the integration of ML-driven predictions into clinical workflows translates into improved patient care.

\section{Conclusion}
\label{sec-conclusion}

In this study, we demonstrated the effectiveness of ML models in predicting survival outcomes for cancer patients using clinical data. XGBoost emerged as the best performing model compared to other models such as Naïve Bayes, Decision Tree, Logistic Regression, and Random Forest. Key predictors such as metastases count, tumor mutation burden, and fraction of genome altered underline the complexity of cancer prognosis and support the integration of clinical and genomic data to enhance patient care. Explainable ML techniques, particularly SHAP, provided a high level of model transparency and interpretability. This is crucial for enabling healthcare professionals and researchers to understand the decision-making processes of the models, supporting more informed clinical decisions and advancing research. The insights from cancer-specific models enriched our findings, offering a detailed look at feature importance across different cancer types. These models are not only accurate and robust but also tailored to meet specific clinical needs. We hope this work inspires further use of advanced ML techniques and comprehensive datasets like MSK-MET to improve the predictive accuracy and clinical utility of models in oncology, ultimately enhancing patient outcomes.


\section{Declarations}
\label{sec-declarations}

\subsection{Ethics approval and consent to participate}
Not applicable.

\subsection{Consent for publication}
Not applicable.

\subsection{Availability of data and materials}
The datasets generated and/or analyzed during the current study are available in the MSK-MET repository at \url{https://www.cbioportal.org/study/summary?id=msk_met_2021}. The code will be available at \url{https://github.com/MU-Data-Science/GAF} after acceptance.

\subsection{Competing interests}
The authors declare that they have no competing interests.

\subsection{Funding}
This work was supported by the National Science Foundation under Grant No. 2201583.

\subsection{Authors' contributions}
P.N. and P.R. designed the study. P.N. implemented and performed the machine learning analysis. P.N. and P.R. analyzed the results. P.N. drafted the initial manuscript. D.R. and P.R. provided feedback and edits for the manuscript. All authors read and approved the final manuscript.

\subsection{Acknowledgements}
This work was supported by the National Science Foundation under Grant No. 2201583. We are grateful to the FABRIC team for their support.

\subsection{Authors' information}
Not applicable.

\bibliographystyle{sn-nature}
\bibliography{references}

\end{document}